\documentclass{article}

\usepackage{arxiv}

\usepackage[utf8]{inputenc} 
\usepackage[T1]{fontenc}    
\usepackage{hyperref}       
\usepackage{url}            
\usepackage{booktabs}       
\usepackage{amsfonts}       
\usepackage{nicefrac}       
\usepackage{microtype}      
\usepackage{lipsum}		
\usepackage{graphicx}
\usepackage{natbib}
\usepackage{doi}

\title{Thermal and quantum fluctuations of harmonic oscillator}

\author{ \href{https://orcid.org/0000-0002-6659-2788}{\includegraphics[scale=0.06]{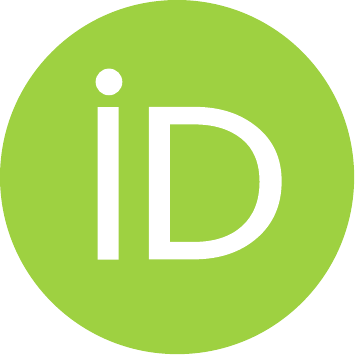}\hspace{1mm}Shigenori Tanaka}\thanks{Corresponding author.} \\
	Graduate School of System Informatics\\
	Kobe University\\
	Kobe, Hyogo 657-8501, Japan \\
	\texttt{tanaka2@kobe-u.ac.jp} \\
\And
	\href{https://orcid.org/0000-0002-9708-6175}{\includegraphics[scale=0.06]{orcid.pdf}\hspace{1mm}Yuto Komeiji} \\
	Health and Medical Research Institute\\
    AIST\\
	Tsukuba, Ibaraki 305-8566, Japan \\
	\texttt{y-komeiji@aist.go.jp} \\
}

\renewcommand{\shorttitle}{\textit{arXiv} Template}

\hypersetup{
pdftitle={A template for the arxiv style},
pdfsubject={q-bio.NC, q-bio.QM},
pdfauthor={David S.~Hippocampus, Elias D.~Striatum},
pdfkeywords={First keyword, Second keyword, More},
}

\begin{document}
\maketitle

\begin{abstract}
We discuss the positional fluctuations of a quantum harmonic oscillator in a heat bath.
Analytic expressions are given for the probability distribution functions of the oscillator position 
in general and limiting (classical and ground state) cases.
\end{abstract}


This is a short note concerning the thermal and quantum fluctuations of a harmonic oscillator. 
This mathematical analysis may be useful for considering the fluctuations of bond lengths involved in molecular dynamics simulations for {\it e.g.} proteins, in which the associated bonding interactions are usually described in terms of (classical) force fields based on the harmonic potentials.
We are mainly concerned with the degree of quantum effects and the validity of classical-mechanical approximations.


Let us consider the Schr\"{o}dinger equation for a particle with the mass $m$ and the coordinate $x$, 

\begin{equation}
\left[-\frac{\hbar^{2}}{2m}\frac{d^2}{dx^2} + U(x)\right]\Psi(x) = E\Psi(x), 
\end{equation}

\noindent
confined in the one-dimensional harmonic potential, 

\begin{equation}
U(x) = \frac{1}{2}kx^2,
\end{equation}

\noindent
where $k$ is the spring constant and $\hbar$ is the Planck constant.
The eigenfunction $\Psi_{n}(x)$ ($n = 0, 1, 2,...$) and the eigenvalue of energy,

\begin{equation}
E_{n} = \hbar\omega\left(n + \frac{1}{2}\right),
\end{equation}

\noindent
are explicitly obtained \citep{Schiff}, where the frequency  $\omega = \sqrt{k/m}$ is introduced.
The quantum-mechanical probability density at the position $x$ is then given by 

\begin{equation}
p_{q}^{(n)}(x) = |\Psi_{n}(x)|^2 = \frac{\alpha}{2^{n}n!\sqrt{\pi}}H_{n}(\alpha x)^{2}\exp(-\alpha^{2}x^{2}),
\end{equation}

\noindent
where $\alpha = \sqrt{m\omega/\hbar}$ and $H_{n}(x)$ refers to the Hermite polynomials \citep{Gradshteyn}.

\par

We here consider the statistical average of the probability density $p_{q}^{(n)}(x)$ over the canonical ensemble at temperature $T$ in the thermal equilibrium.
Recalling the population density of the eigenstate $n$ in the canonical ensemble:   

\begin{equation}
P_{n} = \frac{\exp(-\beta E_{n})}{\sum_{i=0}^{\infty}\exp(-\beta E_{i})} 
= e^{-n\beta\hbar\omega}(1-e^{-\beta\hbar\omega})
\end{equation}

\noindent
with $\beta = 1/k_{B}T$ and $k_{B}$ being the Boltzmann constant, 
we can calculate the statistical probability density at the position $x$ as 

\begin{equation}
P_{T}(x) = \sum_{n}p_{q}^{(n)}(x)P_{n}.
\end{equation}

\noindent
Employing the integral representation for the Hermite polynomials as \citep{Gradshteyn,Tanaka} 

\begin{equation}
H_{n}(x) = \frac{2^n}{\sqrt{\pi}}\int_{-\infty}^{\infty}(x+it)^{n}e^{-t^2}dt, 
\end{equation}

\noindent
we can express Eq.\ (6) as 

\begin{equation}
P_{T}(x) = \frac{\alpha}{\pi^{3/2}}e^{-\alpha^{2}x^{2}}(1-e^{-\beta\hbar\omega})\sum_{n}
\frac{2^n}{n!}e^{-n\beta\hbar\omega}\int_{-\infty}^{\infty}dt(\alpha x+it)^{n}e^{-t^2}
\int_{-\infty}^{\infty}ds(\alpha x+is)^{n}e^{-s^2}.
\end{equation}

\noindent
Then, carrying out the summation over $n$ with 

\begin{equation}
\sum_{n=0}^{\infty}\frac{1}{n!}[2e^{-\beta\hbar\omega}(\alpha x+it)(\alpha x+is)]^{n} 
= \exp[2(\alpha x+it)(\alpha x+is)e^{-\beta\hbar\omega}], 
\end{equation}

\noindent
we find 

\begin{equation}
P_{T}(x) = \frac{\alpha}{\pi^{3/2}}e^{-\alpha^{2}x^{2}}(1-e^{-\beta\hbar\omega})
\int_{-\infty}^{\infty}dt\int_{-\infty}^{\infty}ds\exp[-t^{2}-s^{2}+2(\alpha x+it)(\alpha x+is)e^{-\beta\hbar\omega}].
\end{equation}

\noindent
Finally, performing the Gaussian integrals over the variables $s$ and $t$, we obtain 

\begin{equation}
P_{T}(x) = \frac{\alpha}{\sqrt{\pi}}\sqrt{\tanh\left(\frac{\beta\hbar\omega}{2}\right)}
\exp\left[-\alpha^{2}\tanh\left(\frac{\beta\hbar\omega}{2}\right)x^2\right],
\end{equation}

\noindent
which shows a Gaussian distribution around $x = 0$.
Though this expression itself is known in the literature 
\citep{Messiah,Schonhammer}, the present derivation is very simple and straightforward.

\par

In the limit of zero temperature ($T \to 0$), we see 

\begin{equation}
P_{T}(x) \to P_{T,0}(x)=\frac{\alpha}{\sqrt{\pi}}\exp(-\alpha^{2}x^{2}),
\end{equation}

\noindent
which is the distribution represented by the ground state ($n = 0$).
On the other hand, in the high-temperature (classical) limit ($\beta \to 0$), we find 

\begin{equation}
P_{T}(x) \to P_{T,cl}(x)=\sqrt{\frac{\beta k}{2\pi}}\exp\left(-\frac{\beta}{2}kx^{2}\right), 
\end{equation}

\noindent
which is the Boltzmann distribition with the potential $U(x)$.

\par

It is interesting to compare the above result with the fully classical-mechanical derivation.
Starting with the Newtonian equation of motion,

\begin{equation}
m\frac{d^{2}x}{dt^2} = -\frac{dU}{dx} = -kx,
\end{equation}

\noindent
we find a solution,

\begin{equation}
x(t) = A\cos(\omega t)
\end{equation}

\noindent
with an initial condition of $x(0) = A$ and $\dot{x}(0) = 0$, where $A$ represents the amplitude.
Then, introducing the period $T = 2\pi/\omega$, we can calculate the probability density at the position $x$ as  

\begin{equation}
p_{cl}(x) = \frac{2}{T}\left|\frac{dt}{dx}\right| = \frac{1}{\pi A\sin(\omega t)} = \frac{1}{\pi\sqrt{A^2 - x^2}},
\end{equation}

\noindent
where the amplitude is related to the total energy $E$ of harmonic oscillator as $A = \sqrt{2E/k}$.
It is here remarked that the probability distribution diverges at $x = \pm A$ in the microcanonical distribution 
with a given energy $E$.
Then, transforming from the microcanonical ensemble to the canonical ensemble with a given temperature $T$ in the thermal equilibrium, 
we calculate the statistical probability density at the position $x$ as 

\begin{equation}
P_{T,cl}(x) = \int_{kx^{2}/2}^{\infty}dE e^{-\beta E}p_{cl}(x; E)\left.\right/
\int_{0}^{\infty}dE e^{-\beta E} 
= \frac{\beta}{\pi}\sqrt{\frac{k}{2}}\int_{kx^{2}/2}^{\infty}dE\frac{e^{-\beta E}}{\sqrt{E-kx^{2}/2}}.
\end{equation}

\noindent
The final integration in Eq.\ (17) can be carried out through a change of variables as $E-kx^{2}/2 = z^2$, thus 
leading to the Boltzmann distribution, Eq.\ (13).

\par
  
By using the quantum-mechanical probability density obtained above, we can evaluate 
the fluctuation (variance) of 
particle position around the stable point $x = 0$ as 

\begin{equation}
\langle x^2 \rangle = \int_{-\infty}^{\infty}dx x^2 P_{T}(x) 
= \frac{1}{2\alpha^{2}\tanh\left(\frac{\beta\hbar\omega}{2}\right)} 
= \frac{\hbar}{2m\omega}\coth\left(\frac{\beta\hbar\omega}{2}\right).
\end{equation}

\noindent
Due to $\tanh\left(\frac{\beta\hbar\omega}{2}\right) \to 1$ for $\beta \to \infty$,  we find  

\begin{equation}
\langle x^2 \rangle \to \langle x^2 \rangle_{0} = \frac{1}{2\alpha^2} = \frac{\hbar}{2m\omega}
\end{equation}

\noindent
in the zero temperature limit. 
Recalling $\tanh\left(\frac{\beta\hbar\omega}{2}\right) \to \frac{\beta\hbar\omega}{2}$ in the classical (high-temperature) 
limit, on the other hand, we find 

\begin{equation}
\langle x^2 \rangle \to \langle x^2 \rangle_{cl} = \frac{1}{\alpha^{2}\beta\hbar\omega} 
= \frac{k_{B}T}{m\omega^2}.
\end{equation}

\noindent
It is noted that we see 
$\langle x^2 \rangle \ge \langle x^2 \rangle_{cl}$ because of 
$0 \le \tanh\left(\frac{\beta\hbar\omega}{2}\right) \le \frac{\beta\hbar\omega}{2}$.
We also see 
$\langle x^2 \rangle \ge \langle x^2 \rangle_{0}$
owing to 
$\tanh\left(\frac{\beta\hbar\omega}{2}\right) \le 1$.
Thus, the evaluation of $\langle x^2 \rangle$ in Eq.\ (18) is given as a combination of quantum and thermal contributions.
Since we observe 

\begin {equation}
\frac{\langle x^2 \rangle_{cl}}{\langle x^2 \rangle_{0}} = \frac{2k_{B}T}{\hbar\omega}, 
\end{equation}

\noindent
the contributions from the thermal (classical) and quantum fluctuations are dominant in the high-temperature 
(or low-frequency) and low-temperature (or high-frequency) regions, respectively.


\section*{Acknowledgement}

\noindent
S.T. would like to acknowledge the Grants-in-Aid for Scientific Research (No.\ 21K06098) from the Ministry of Education, Cultute, Sports, Science and Technology (MEXT).





\bibliographystyle{unsrtnat}
\bibliography{references}  






\end{document}